\documentstyle[epsfig,aps,twocolumn]{revtex}

\begin{document}
\draft
\preprint{ }

\title{First excitations of the spin 1/2 Heisenberg antiferromagnet on the
\protect\textit {kagom\'e} lattice}
\author{
Ch. Waldtmann
\thanks{Institut f\"ur Theoretische Physik,
Universit\"at Hannover, D-30167 Hannover, Germany},  H.-U. Everts $^* $,
B. Bernu
\thanks{Laboratoire de Physique Th\'eorique des
    Liquides, Universit\'e P. et M. Curie, case 121, 4 Place Jussieu, 75252
    Paris Cedex. URA 765 of CNRS},
P. Sindzingre $^{\dag} $,
C. Lhuillier $^{\dag} $.
P. Lecheminant 
\thanks{Groupe de Physique Statistique, Universit\'e de Cergy-Pontoise,
 95302 Cergy-Pontoise Cedex}
L. Pierre 
\thanks{Universit\'e Paris-X, 92001 Nanterre Cedex}}
\maketitle

\abstract{ 
We {\rm study} the exact low energy spectra of the spin $1/2$ Heisenberg 
antiferromagnet on small samples of the {\it kagom\'e} lattice of up to
$N=36$ sites.
In agreement with the conclusions of previous authors, we find that
 these low energy spectra 
contradict the hypothesis of N\'eel type long range order. Certainly,
the ground state of this system is a spin liquid, but its properties
are rather unusual. The magnetic ($\Delta S=1$) excitations are
separated from the ground state by a gap. However, this gap is filled
with nonmagnetic ($\Delta S=0$) excitations. In the thermodynamic limit
the spectrum of these nonmagnetic excitations will presumably develop
into a gapless continuum adjacent to the ground state.
 Surprisingly,
the eigenstates of samples with an odd number of sites, i.e. samples with
an unsaturated spin,  exhibit
symmetries which could support long range chiral order. We do not know if these
states will be true thermodynamic states or only metastable ones. In any case,
the low energy properties of the spin $1/2$ Heisenberg 
antiferromagnet on the {\it kagom\'e} lattice clearly distinguish this system
from either a short range RVB spin liquid or a standard chiral spin liquid. 
Presumably they are facets of a generically new state of frustrated
two-dimensional quantum antiferromagnets.
} 
\section{Introduction}
        Zeng and Elser \cite{ze90} were the first to point out that the
spin $1/2$ Heisenberg antiferromagnet on a {\it kagom\'e} lattice 
(hereafter called KHA) could be disordered.
Since then  many studies have been devoted to this
system. {\rm A}ll quantum approaches based  on {\rm either} exact
diagonalizations \cite{ce92,le93}, on {\rm perturbational} series 
expansions \cite{sh92}  or on high
temperature {\rm expansions}\cite{ey94} point to a disordered ground state.
{\rm Previous exact diagonalization studies \cite{ce92,le93} have shown} that
spin-spin, spin nematic, spin Peierls and
chiral-chiral correlations decrease very rapidly with distance. {\rm
Therefore 
 the idea that the KHA is a spin liquid is by now widely accepted.}
The precise nature of the spin liquid state of the KHA and of its low
lying excitations is, however far from clear. Some approaches point to
a liquid of short range dimers \cite{e89,s92,ze95}, others suggest
that the KHA might be a candidate for a chiral spin liquid state of
the kind first proposed for the triangular antiferromagnet by
Kalmeyer and Laughlin \cite{kl87,mz91a,ywg93}.
It seems to be the common view that in both pictures any excitations,
magnetic or nonmagnetic ones, are separated from the ground state by a
finite gap.
While for the magnetic excitations this view is supported by all previous
numerical work on the KHA \cite{ze90,le93}, the situation is less
clear for the nonmagnetic excitations. The only piece of evidence for a finite albeit
small gap separating the singlet excitations of the KHA from its
ground state comes from the work
of Zeng and Elser \cite{ze95}. Elaborating on the idea of a
short-range dimer liquid these authors construct an effective
low energy hamiltonian for the singlet subspace of the KHA, 
which as expected, exhibits a small gap above its ground state.
The question of whether one of these pictures, that of a 
short range RVB spin liquid or
that of a  chiral spin liquid, applies to the KHA has led us to
 study  not only the energies but also the 
symmetry properties of a very large number of low lying levels of
the exact spectra of the KHA.

\section{Numerical approach and some general results}
Using {\rm the} Lanczos technique and  {\rm a} complete group theoretical  
analysis (as in \cite{blp92}) we 
computed the low energy spectra of the Heisenberg hamiltonian  
\begin{equation}
{\cal H} = J \sum_{<ij>} 2 {\bf S}_i. {\bf S}_j
\end{equation}
for small samples of the {\it kagom\'e} lattice on a 2-dimensional torus 
with periodic and twisted boundary conditions
(in the following $J=1$).
We have obtained a very large number of low lying levels  
in each irreducible representation (IR) of $SU(2)$ and of the space
groups of the  $ N= 9, 12,
15, 18, 21, 24, 27,36$ samples.

As in all spectra of Heisenberg antiferromagnets that we have examined,
the ground state energy of the $S$-subspaces increases with the total
spin $S$.
However, the spectra of the finite samples of the KAH do not show the
pattern that is characteristic for systems  which exhibit N\'eel type long
range order in the thermodynamic limit as, e.g.  the triangular
antiferromagnet (for a detailed discussion of the signature of long
range order in the spectra of finite samples see ref.
\cite{bllp94}).
Instead, all the signatures of a ``liquid'' are present in the low 
energy spectra of the KAH: {\it i)} The lowest levels associated 
with the different momenta {\bf k} in the Brillouin zones of the
samples  are almost independent of  ~{\bf k}. For instance, for the largest 
sample $(N=36)$, the energies of these lowest levels vary by less 
than $0.2\%$ when {\bf k} is varied through the four inequivalent 
{\bf k}-points of the Brillouin zone (see Table 1). This absence of dispersion 
excludes the possibility of a broken translational symmetry. 
{\it ii)} The system is extremely soft against any twist of the
boundary 
conditions \cite{lblps97}
. {\it iii)} The size dependence of the ground state energy 
is more than one order of magnitude smaller than for the triangular 
antiferromagnet in the same range $9\le N\le 36$.      

\section{Magnetic gap and nonmagnetic excitations}
Our results, (Fig.~1), confirm the conclusion of Elser and coworkers \cite{ze90,le93} about 
the existence of a gap for the magnetic
$ \Delta S =1$ excitations.
 The thermodynamic limit of this gap 
cannot be extracted from the present data
with high accuracy. The extrapolations shown in Fig. 1  point to a lower bound  of
about $J/20$ (see also the discussion in ref. \cite{lblps97}).
In any case, the value of the magnetic gap appears to be at least one
order of magnitude smaller than the exchange 
energy $J$ which is needed to break an isolated singlet pair. This is
a first indication  that the picture of the {\it kagom\'e} antiferromagnet as a spin liquid consisting of short-range 
singlet dimer pairs may be inappropriate. Nevertheless, it comes as a
surprise that quite contrary to the standard picture of a
spin liquid the magnetic gap of
the KHA is filled with nonmagnetic excitations. This is illustrated in
Fig.~2, where we display the integrated density of states of the
$N=36$ sample. In this sample there are $183$ singlet levels below the
lowest triplet. Similarly, in the $N=27$ the $\Delta S=1$ gap is
filled with $153$ nonmagnetic excitations.
 For the samples we have examined, the number of
nonmagnetic ($\Delta S=0$) excitations
within the magnetic gap
grows roughly as $\alpha^N$ with the system size, where $\alpha\simeq
1.15$ and $\alpha\simeq 1.18$ for the even and odd samples,
respectively. These results call for a few comments:

 {\it i)} The density of nonmagnetic levels above the ground states of
the two largest samples ($N=27, 36$) strongly suggests that in the thermodynamic
 limit the nonmagnetic excitation spectrum of the KAH is a {\it gapless
continuum adjacent to the ground state}. Interestingly, an algebraic
decay of certain correlation functions would be compatible with such a
gapless nonmagnetic continuum. A candidate is the dimer-dimer correlation function,
\begin{equation} C_{<i,j><k,l>}=<({\bf S}_i\cdot {\bf S}_j)({\bf S}_k\cdot {\bf
S}_l)>-<{\bf S}_i\cdot {\bf S}_j>^2, \end{equation} which might decrease
algebraically with the distance between the
nearest neighbor bonds $<i,j>$ and $<k,l>$. Leung and Elser
\cite{le93} have calculated this function for the
ground state of the $N=36$ sample of the KHA. Their result, a damped
oscillatory decrease of $C_{<i,j><k,l>}$,
is not inconsistent with an algebraic decay.
An algebraic decay of the
dimer-dimer correlations would imply that the system is critical with
respect to nonmagnetic quantum fluctuations. Whether this is the case
for the KHA or not will be impossible to decide from finite size
studies of the correlation function alone.

{\it ii)} The exponentially large number of low lying singlets of the
spin $1/2$ KHA is reminiscent of the ground state degeneracy of the
corresponding classical model. Stability considerations show that only
planar spin configurations qualify as true classical ground states
\cite{s92,c92,chs92,ccr93}. Their number and hence the
ground state degeneracy of the classical KHA grows as $1.134^N$ with
the system size $N$ \cite{hr92}. The different planar configurations
are connected with each other  by local rotations in spin
space \cite{chs92}. In a semiclassical picture these local rotations
provide tunneling paths between the different planar configurations.
 It is therefore tempting to think of the
low lying singlets of the quantum KHA as of tunnel-split classical
ground states. This semiclassical picture has been pursued by von Delft and
Henley \cite{vdh93}. As one of their main results these authors find
that for spin $1/2$ all the tunneling events involving the coherent
rotation of the spins on simple loops on the {\it kagom\'e} lattice  yield zero tunneling
amplitudes due to destructive interferences between different
tunneling paths. The smallest effective tunneling event requires the
coherent rotation of the spins on two nested hexagons on the
{\it kagom\'e} lattice involving $24$ spins in total. Certainly, this
scenario cannot explain the abundance of low lying singlets in a
sample consisting of only $21$, $27$, or $36$ spins. In our view, this high density
of low lying singlets in small samples of the KHA is a strong argument
against the validity of the semiclassical approach for the spin $1/2$ KAH.

{\it iii)} Another comparison that suggests itself is between the density of
low lying singlets and the dimension of the nearest neighbor
valence-bond basis on the {\it kagom\'e} lattice which grows as
$2^{N/3}=1.26^N$ with the system size \cite{e89}. From this point of view the variational approach of Zeng
and Elser \cite{ze95} which builds on the valence bond basis seems
fully justified.    
 The main difference
between the approximate singlet spectrum  of Zeng and Elser \cite{ze95} and our
numerically exact result lies in the density of levels at the bottom
of this spectrum (see Fig.~2). The absence of a gap above the ground state and the high density of very low lying levels in the
exact spectrum suggests that a dimer product representation of the
corresponding eigenstates will necessarily  contain long range singlet pairs.
The importance of such longer range singlets in the spin liquid
picture of the KHA might have been anticipated from the work of Zeng and Elser who
observed that the inclusion of first {\it and}
second-neighbor singlet pairs in their variational Hilbert space led
to a considerable improvement over the results obtained in a pure
first-neighbor dimer basis. 

{\it iv)}
The exponential number of non magnetic
states in the gap should be visible in various experimental situations:
It may explain the vanishing of the neutron elastic forward scattering
cross section \cite{lbar96}  and  
the very weak field dependence of the low temperature heat capacity of $SrCrGaO$
observed by Ramirez and coll. \cite{r97}. Specific heat data and
neutron scattering experiments both point to a density of states
that increases linearly with the energy. In the very low energy region
($E/N=-0.4384 \dots -0.4364$) our 
numerical data are consistent with such a  linear energy dependence of
the density of states (see the inset in Fig. 2, where the integrated density 
of states is plotted); they are definitely
not consistent with an energy independent density of states as it
would be  obtained for an ordinary spin glass \cite{r94}.

\section{Spin 1/2 excitations}
\subsection{Symmetries of the low lying S=1/2 eigenstates.}
The second surprise lies in the 
 degeneracies of the  low lying levels of the 
spectra of the samples with an odd number of
spins ($N=9,21,27$) and in the symmetries of the eigenstates
associated with them.
We are focusing here
on the exponentially large number of spin doublets
which fill the gap between the ground state and the lowest  $S=3/2$ eigenstate.
On account of the finite size scaling of the  available spectra,
we can ascertain
that  in the thermodynamic limit 
 all these eigenstates
are  complex and exhibit an at least twofold degeneracy in addition to
their magnetic degeneracy. (In the following we shall
consistently ignore the trivial spin degeneracy.)
 Hence all these
states may participate in a spontaneous  mechanism breaking the
time reversal 
(and parity) invariance of the Heisenberg hamiltonian.
In the spectra of the smallest samples, there are still some  low lying
 levels which belong to 1D IRs of
the space
group and are hence nondegenerate, but when the size of the
sample  increases these levels are pushed towards higher energies:
in the $N=27$  sample there are 153 $S=1/2$ levels below the 
lowest $S=3/2$
level, only $3$ of them belong to  1D IRs of the space group.
The finite size
extrapolation predicts that for  odd sizes $N >  50$ no states
from  the  1D IRs 
of the space group will be left in the magnetic gap.
This is indeed a very specific property of the KAH which we have never
observed in the spectrum of any spin hamiltonian on the triangular lattice 
: in the usual triangular N\'eel antiferromagnet the homogeneous 
low lying states
belong to the trivial representation of the $C_{3}$ group, whereas in
the present case the $S=1/2$ levels of the trivial representations are pushed high
up in the spectra.\footnote{ In the case of the $q=0$ N\'eel order,
 enforced on the {\it kagom\'e}
lattice by a  large enough second neighbor antiferromagnetic  coupling
\cite{we97}, the
ground state of the odd samples does in fact belong to the
two complex conjugate IRs of $C_{3}$ \cite{we97}.
However in this case it is the only low lying
state with this symmetry,
 all the others have much higher energies.}
This observation leads us to concentrate on possible new  properties
of this low lying $S=1/2$ continuum.

\subsection{ Chiral  observable and correlation function}

All these low lying levels (except the 3 belonging to 1D IRs)
can sustain non-zero
expectation values of the chirality operator
\begin{equation}
\Xi_{123} = 2 \: {\bf
S}_{1}\cdot ({\bf S}_{2} \times {\bf S}_{3})= Im(P_{123})
\end{equation}
where $P_{123}$ is the cyclic permutation  of three spins
around a triangular plaquette.

We have measured this quantity in the lowest lying states of the
$N=9,21$ and $27$ samples 
for triangles with  side lengths $1$, $\sqrt{3}$, $2$,
$\sqrt{7}$, $3$
 (see Table 2).
Compared with the eigenvalues $\pm \sqrt{3}/2$ of the operator
$\Xi_{123}$ in the $S=1/2$ bound states of three spins,
the measured expectation values are indeed
quite small, and there
is a sharp decrease from the $N=9$ to the $N=21$ sample. But
in view of the data for $N=21$ and $N=27$ it is unclear how
this observable will behave for larger sizes. The chiral-chiral
correlation function  ( last lines of Table 2)
is indeed short ranged, which may explain
the quasi-absence of size effects between $N=21$ and $N=27$,
and is indeed characteristic of a liquid.

In order to have an indication of the possible order of
magnitude of the chirality in a pure chiral spin
liquid, we have computed the expectation value of $\Xi_{123}$
for the spin $1/2$ excitations of the Laughlin wave function
\cite{ywg93} on the same small lattices; these expectation
values  are   denoted (L.w-f) in Table 2. 
While for triangles of size 1 the L.w-f values are an order
of magnitude larger than the exact results, they are
of the same order of magnitude for larger sizes.
By this comparison we do not want to suggest that the
physics of the KAH is the same as that of the Laughlin
picture of a chiral spin liquid ( as has been found in section 4. above, 
the  first excitations are different).
We only want to point out that the exact results may be 
significant in spite of their numerically small values.

On the basis of results obtained by applying the $SU(2)$ Schwinger-boson
approximation to samples of the KAH with sizes of up to $N=72$ sites Sachdev
argues
that the chirality of the ground state goes to zero with increasing 
system size \cite{s92}. It is indeed true that the  exact
ground states of the even $N$ samples have zero chirality for 
$N=12,18,24,36$. The present work indicates that this may not
be true for the odd $N$ samples.

\subsection{Chern index of the low lying  S=1/2 levels}
Searching for a quantity that might provide further evidence
for or against collective chiral behavior we followed the proposal of
Haldane and Arovas \cite{ha95} and computed the Chern number of
the homogeneous (${\bf k}= {\bf 0}$) ground state of the $N=9$
sample. 
It is a state which transforms as ${\it R_{2\pi/3}\psi =
e^{i2\pi/3}}\psi$ under a spatial $2\pi /3$
rotation. In this entire  paragraph we shall concentrate on these
specific states, which are the low lying homogeneous states of
the $S=1/2$ continuum that we have been  discussing above.
Anticipating further results we shall refer to these eigenstates
as ``chiral'' states henceforth.
In the $N=27$ sample, $36$ of the $39$ homogeneous states
of this low lying continuum are chiral states,  and we expect that all
states of this continuum will  be chiral in the thermodynamic limit.
The Chern number is a topological property
of the eigenfunctions of a system constrained by twisted boundary conditions
\cite{ha95}. If ${\bf T}_1$ and ${\bf T}_2$ denote the two vectors
defining the sample cell, the twisted boundary conditions are defined
by
\begin{equation}
 S^{\pm}( {\bf R}_{i}+{\bf T}_{j}) = e^{ \pm i \Phi_{j}}
S^{\pm}({\bf R}_{i}),
\end{equation}
where $\Phi_{1}$ and $\Phi_{2}$ are the two angles defining
the rotations of the spins around the $z$ direction.
Let us denote {\rm by} $|n(\Phi_{1},\Phi_{2})>$
the eigenstates obtained by adiabatic evolution of the state $|n>$  under the 
twists. The Chern index of the ket $|n>$ is defined by  
\begin{equation}
\label{eq-chern}
 C(n) =  \frac{i}{2 \pi} \int_{0}^{2 \pi}
\!\!\!\!d \Phi_{1}\int_{0}^{2 \pi}\!\!\!\!d \Phi_{2} \;\;
\varepsilon_{ab} < \frac {\partial n}
{\partial \Phi_{a}}  | \frac {\partial n} {\partial \Phi_{b}}>.
\end{equation}
It is a topological constant that can take only integer values. It can
only be non-zero for complex, degenerate 
 eigenstates, and, as will be discussed
below, a non-zero value of the Chern number of an eigenstate $|n>$
implies 
peculiar  physical properties of the system in this state.   
A formula equivalent to (4) is obtained by the use of Stokes'
theorem:
\begin{equation}
C(n)=2\pi i\oint <n|dn>.
\end{equation} 
It shows that the Chern number counts the number and the nature of the
singularities of the phase of $|n(\Phi_1 , \Phi_2) >$ in the ($\Phi_1$,
$\Phi_2 $) Brillouin zone. It is a measure of the vorticity of this
phase. 
For computational purposes the expression (4) is quite cumbersome.
A more convenient expression is obtained by changing to a spatially
varying reference frame for the spin variables  such that
 the twisted boundary conditions (3) are replaced by periodic ones.
The main steps in this procedure are as follows: Let ${\cal S}_{0}$ be the
reference frame for the spin at the origin ${\bf R}_0$ of the sample.
Then the reference frame at the site ${\bf R}_i$ is obtained from
${\cal S}_0$ by a rotation through the angle 
\begin{equation}
        \Theta( {\bf R}_i) = ({\bf R}_i -{\bf R}_0)({\bf e}_1 \theta_1 + {\bf e}_2
        \theta_2),
\end{equation}
around the $z$-direction, where  ${\bf e}_1$,  ${\bf e}_2$ span the
Bravais lattice and 
$\theta_1$, $\theta_2$ are the increments of the rotation angle along
${\bf e}_1$ and  ${\bf e}_2$. $\theta_1$, $\theta_2$ are chosen such that they add up to
the twist angles, i.e.:
\begin{equation}
\Phi_a=\Theta( {\bf R}_0 + {\bf T}_a ) ,\qquad a=1,2.
\end{equation}
In this spatially varying reference frame the hamiltonian reads
\begin{equation}
   \tilde{{\cal H}}(\theta_1,\theta_2)=2\sum_{<i,j>} \{ S_i^z S_j^z 
   + 1/2 \left( e^{i\chi_{ij}}S_i^+ S_j^-  + h.c.\right) \},
\end{equation}
where
\begin{equation}
 \chi_{ij} = \Theta({\bf R}_i)-\Theta({\bf R}_j).
\end{equation}   
The Chern number in the eigenlevel $|n>$ associated with the
eigenvalue $\tilde{E}_n$ of $\tilde{{\cal H}}$ is then readily obtained as the average over the
($\Phi_1$, $\Phi_2$) Brillouin zone of the function :\\
\begin{equation}
	{\sf{K}}_{ab}= 2\pi \epsilon_{ab}\sum_{p\ne n}
        \frac{<n|\frac{\partial \tilde{{\cal H}}}{\partial \Phi_a}|p><p|\frac{\partial
	 \tilde{{\cal H}}}{\partial
\Phi_b}|n>}{(\tilde{E}_n-\tilde{E}_p)^2},
\end{equation}

where

\begin{equation}
\frac{\partial{\tilde{{\cal H}}}}{\partial \Phi_a} = i \sum_{i,j}
         \left( e^{i\chi_{ij}}S_i^+ S_j^-  - h.c.\right) 
       \frac{\partial \chi_{ij}}{\partial \Phi_a}.
\end{equation}
The phase angle $\chi_{ij}$  depends linearly on
$\Phi_{a}$ through the local  twists (7), and $\frac{\partial \chi_{ij}}{\partial \Phi_{a}}$ is a
linear combination of the space components of $(\bf{R}_i-\bf{R}_j)$.

The computation of the Chern number is a heavy task.
After a thorough study of the chiral ground state of the $N=9$ sample we can
ascertain that for this state the Chern number is $+1$
(respectively $-1$ {\rm for} the complex conjugate eigenstate).
In view of our study of the singular points of 
$\tilde{E}_{n}( \Phi_{1},\Phi_{2})$ in the chiral states of the $N=21$ and $27$
samples,
we expect it to be odd {\rm for} any of the low lying doublets 
and most probably equal to
$\pm 1$  for the ground state (in this last case, we find only one
conical point at  $(\Phi_{1},\Phi_{2}) = (0,0)$).

The physical significance of
the Chern index may be inferred from the expression (11) which is a zero frequency
Kubo response function. The general formula for the response of
an observable B to an excitation of the observable A
reads at $T=0$:
\begin{equation}
\chi^{T=0}_{BA}(\omega = 0) = 2\hbar \; Im\sum_{p\neq
n}\frac{<n|B|p><p|\dot{A}|n>}
{(\tilde{E}_n-\tilde{E}_p)^2},
\end{equation}
where $\dot{A}$ is the time derivative
of the observable coupled to
the external field. Comparing this general form of the response
function with the special form (11), one sees that in (11) both
operators $\dot{A}$ and $B$ are linearly connected with the total spin
currents $J_a=i\sum_{<i,j>}\left( e^{\chi_{ij}} S^+_i S^-_j
-h.c.\right)$, where the $\sum_{<i,j>}$ is to be restricted for $a=1,2$,
to the bonds  along the directions ${\bf e}_1$ and ${\bf e}_2$.
The external field that drives a spin current along, e.g. the
${\bf e}_1$ direction, is a magnetic field with a constant gradient in
this same direction. This is most easily seen on the example of the
square lattice, where the relation between $(\theta_1,\theta_2)$ and 
$(\Phi_1,\Phi_2)$ is diagonal. The case of the {\it kagom\'e} lattice is
technically slightly more involved, but the physics is the same. The
perturbation of the hamiltonian $\tilde{\cal{H}}$ induced by a
magnetic field $B^z$ with a constant gradient in the ${\bf e}_1$
direction is given by 
\begin{equation}
V=-\sum_i \alpha_1 X_{i1} S_i^z,   
\end{equation}

where 
$X_{i1}=({\bf R}_i\cdot {\bf e}_1)$.
Here, $\alpha_1$ is the linear  increment  of the magnetic field over
one lattice constant. (Here and in the sequel we set $\hbar$, the Bohr
magneton and the gyromagnetic factor equal to unity). Identifying the
observable $A$ as  
$A=\sum({\bf R}_i \cdot {\bf e}_1)S_i^z$ one sees immediately that its time
derivative is the total spin current in the $\bf{e_1}$ direction:
\begin{equation}
\dot{A} =-i \left[A,H \right]=  \sum_{<i,j>}Im\{ e^{ i
\chi_{ij}}S_i^+ S_j^-  \}(X_{i1}-X_{j1})  = J_1
\end{equation}
In the orthogonal geometry one has the
simple relation $J_a= \frac{\partial H}{\partial \Phi_a}$,
 $a= 1,2$, and
the expression (11) takes the form of a transverse current-current correlation
function:
\begin{eqnarray}
{\sf{K}}_{ab}= 2 \;Im\sum_{p\ne n}\frac{<n|J_a|p><p|J_b|n>}{(\tilde{E}_n-\tilde{E}_p)^2}
\end{eqnarray}
This correlation function measures the transverse current $J_2$
generated in the ${\bf e}_2$ direction by a gradient of the magnetic
field $B^z$ along the ${\bf e}_1$ direction. Hence we arrive at the
linear relation
\begin{eqnarray}
<J_2>= 2\pi{\sf{K}}_{ab}\frac{ \partial B^z}{ \partial X_1}
\end{eqnarray}
At this point it is important to note that the above derivation of
(17) is purely formal. It is a linear expansion with respect to the
perturbation $V$, (14), which, as it is obvious from (14), becomes
arbitrarily large in the thermodynamic limit. Thus, (17) cannot be
considered a physically meaningful relation between the spin current
and the gradient of the inhomogeneous magnetic field. However, by
employing a gauge argument Haldane and Arovas \cite{ha95} show that a
physically sound relation is obtained from (17), if the response
function ${\sf{K}}$ is replaced by its average over the torus
$0<\Phi_1,\Phi_2<2\pi$ of the twist angles, i.e.~by the Chern number
$C(0)$.

 From the mathematical point of view the Chern number of this
spin system is equivalent to the TKNN index of the  Quantum Hall
Effect (QHE)\cite{tknn82}, and as in the  QHE the spin $1/2$ excitations
are separated from the $S=0$ ground state by a finite gap (see
 $\Delta_{S=1/2}$ in
Fig.3). But there are very definite and crucial differences:
{\it i)} In our spin system, contrary to the QHE, the parity
and time reversal invariance are not externally broken. Eigenstates with
positive and negative Chern number are degenerate.
{\it ii)} In each spin sector there is a continuum of
excitations adjacent to the ground state and there may be
couplings between the two sectors under the effect of an
external magnetic field. So it is difficult to imagine how the
microscopic rigidity associated with the Chern number could become manifest on a macroscopic level.
 {\it iii)} Finally our numerical results seem to indicate that the creation of two
spin $1/2$ excitations is {\it less} favorable than the creation of a spin $1$
excitation : $2 \, \raise 2pt\hbox{{\Large .}} \mbox{E}_{\Delta{S=1/2}} > 
\mbox{E}_{\Delta{S=1}} $ (see ~Fig.~3),
but because of the uncertainties in the extrapolation
procedure, this last result is not entirely reliable.
The question of whether the excitation of two spin $1/2$ entities is
energetically more or less favorable than the creation of one $S=1$ excitation 
 is indeed an  major open point.
In the first case the spin $1/2$ excitations
would be true thermodynamic excitations, in the opposite case
they could only appear as metastable excitations in
sophisticated dynamic experiments.
Whatever the answer to this last question may be, the KAH is certainly
not a system which fits easily into the frame of standard
continuum chiral theories.
In any case the symmetries of these spin $1/2$ excitations
are features 
that will survive in the thermodynamic limit although the states may 
be metastable in this limit.
We may hypothesize that their  non-zero Chern number, since it is a quantum
number, will also survive in the thermodynamic limit.
In this picture the chiral $S=1/2$ states of the KAH can
certainly not be viewed as
simple bound states of three spins $1/2$ in a sea of singlets:
a cluster of three spins in its $S=1/2$ ground state has indeed
a nonzero expectation value of the chirality $\Xi_{123}=\pm \sqrt{3}/2$, 
but its  Chern index is zero. The Chern number is a measure of a topological
rigidity of the 
N-particle states,  which is by itself unique. To our knowledge the KAH
 is the first system with a hamiltonian that does not break
parity and time reversal invariance, where a non-zero Chern
number has been observed.

\section{Conclusion}
In conclusion, our numerical study of the low lying spectra of the
spin $1/2\,$ KAH leads us to assert that this system is a strange
spin liquid:
Its lowest excitations are  soft nonmagnetic excitations whose 
existence in a spin system has not been
anticipated in previous theoretical investigations of antiferromagnets. The
absence of a
gap between these excitations and the ground state may signal that the
KAH is, at $T=0$, in a critical state with arbitrarily long-ranged
singlet-singlet correlations. 
In any case,
the existence of this continuum of low energy nonmagnetic modes
in the KAH already contradicts the conception of the ground state of
this system as a short-range RVB state.
It  also  contradicts the classical and semi-
classical pictures of the KAH which are all more or less based on the
idea of preferred planar spin arrangements, which imply a breaking
of the $SU(2)$ symmetry. As has been shown in previous work
\cite{blp92,bllp94}, the spectra of {\it finite} samples of a spin system have to exhibit a
certain signature, if a breaking of the $SU(2)$ symmetry is to occur in the {\it
thermodynamic limit}.
The spectra of the finite
samples of
the KAH do {\it not} show this signature \cite{lblps97}, and thus a planar
arrangement of the spins of the KAH can be ruled out.

In the sea of  long-range correlated singlets
 an unpaired spin $1/2$ is surrounded by transverse spin currents,
which are another signature of  long range correlations.
The non zero Chern number of the $S=1/2$ states of the KAH
is a proof of some kind of microscopic rigidity (sensitivity
to the boundary conditions), which is
usually found in chiral theories in the literature.
However, in standard chiral theories these spin $1/2$ excitations
are indeed the lowest excitations of the system, which is definitely not the
case for the KAH.
We do not know if these spin $1/2$ excitations
will be true thermodynamic excitations or only metastable
ones,
but they might show up in dynamical experiments (transverse spin diffusion, spin echoes).

To conclude we want to emphasize that there is some 
evidence that this kind of ground state and low
lying excitations ( critical  singlets and gapped magnetic
excitations) are  robust properties of the KAH: 
 finite perturbations (second neighbor interactions) are needed to drive
the KAH into N\'eel ordered states \cite{lblps97,we97}, and the XY model on
the {\it kagom\'e} lattice has a spectrum  that is similar to the spectrum 
of the Heisenberg
model. Thus, the above properties of the KAH with nearest neighbor
interactions may be facets of a new state of frustrated quantum
antiferromagnets which is generically different from the N\'eel ordered state on
the one hand and from the short-range RVB state on the other \cite{sz92,tnyk95,khsk97,tku96,am96a,am96b}.
 The low energy physics of the KAH may be embodied
in real compounds:
In spite of the fact that $SrCrGaO$ is a spin $3/2$
 compound with many complexities (dilute triangular planes of spins
intercalated between the {\it kagom\'e} planes,
 role of the defects), some of its unusual properties
(low temperature specific heat, dynamical spin correlations)
may be explicable on the basis of the present results.

$\bf{Acknowledgements}$:
This work started from discussions
in Les Houches with  P. Chandra, and P. Coleman
 (Session on ``Strongly interacting fermions and high
$T_{c}$ superconductivity'' in July 91) and has benefited from very
useful interactions in Aspen 
(Session on ``Quantum antiferromagnetism'',
 July 96). We thank A. P. Ramirez for communications of his results
before publication. 
We   acknowledge very interesting exchanges with P. Azaria.
Computations were performed on a C90 at the Institut de D\'eveloppement des
Recherches en Informatique Scientifique
of C.N.R.S. under contracts 940076-960076, on a CRAY T3D at the 
Konrad-Zuse-Zentrum f\"ur Informationstechnik, Berlin, and on the CRAY T3E-512
of the Zentralinstitut f\"ur Angewandte Mathematik, Forschungszentrum J\"ulich.


\begin{figure}[h]
\begin{center}
\unitlength1cm
\begin{picture}(9,9)
\put(0.0,-.5){\epsfig{file=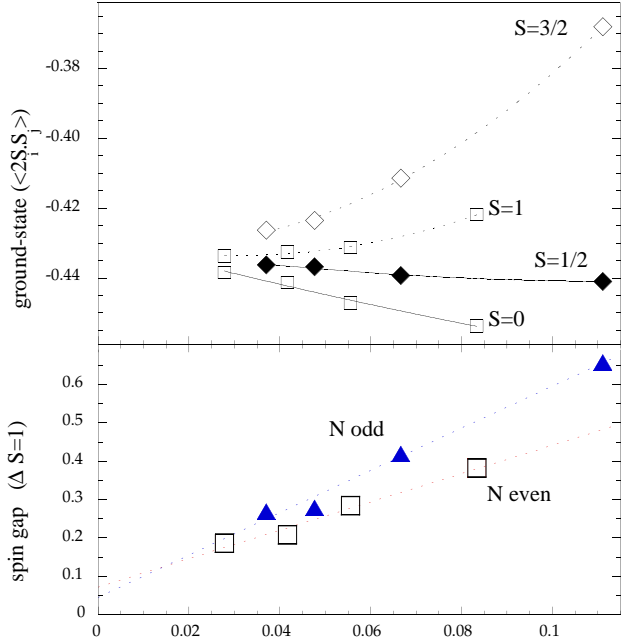,height=9cm,angle=0}}
\end{picture}
\end{center}
\caption{a)
Energy per bond 
$\left<2{\bf S}_{i}.{\bf S}_{j}\right>$ 
in the ground states of $S_{min}$ and $S_{min}+1$.
 The
energy per bond in the thermodynamic limit is approximately
$ 2<{\bf S}_{i}.{\bf S}_{j}> = -0.43$. b) Spin
gap versus $1/N$ for the even $N$ and the odd $N$ samples.}
\label{fig.1}
\end{figure}

\begin{figure}[h]

\begin{center}
\unitlength1cm
\begin{picture}(7,7)
\put(-1.1,6.9){\epsfig{file=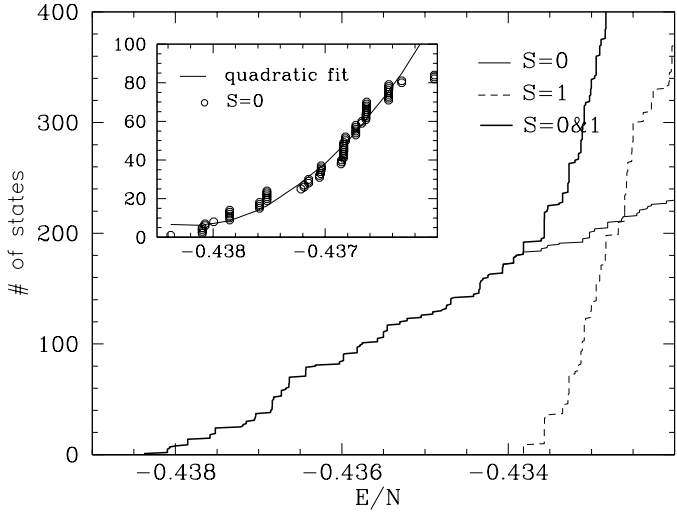,height=9.4cm,angle=-90}}
\end{picture}
\end{center}
\caption{
Integrated density of states of the low lying levels of the $N=36$
sample plotted versus the energy per bond (heavy line).
The light and the broken line display the same quantity for
the $S=0$ and $S=1$ subspaces separately.
The inset shows a quadratic fit to the lowest 70 states of the spectrum.}
\label{fig.2}
\end{figure}

%

%
\begin{figure}[h]

\begin{center}
\unitlength1cm
\begin{picture}(8,8)
\put(-0.5,-0.0){\epsfig{file=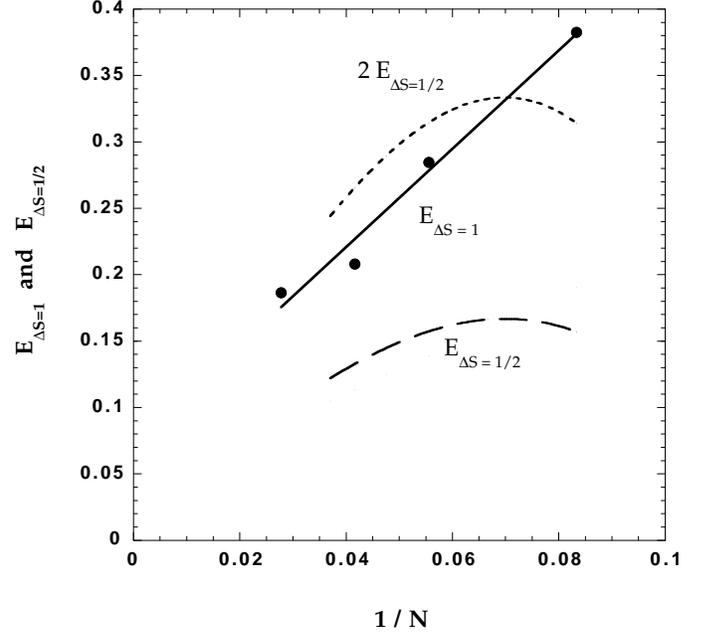,height=8.5cm,angle=0}}
\end{picture}
\end{center}
\caption{ 
$E_{\Delta S=1/2}$ : energy for creating one $S=1/2$ excitation. 
This energy is the difference between 
the interpolated ground state 
energies for $S=1/2$ and $S=0$ (see Fig.1).
$E_{\Delta S=1}$ : energy for creating one $S=1$ excitation in the 
even samples (magnetic gap). The graphs shown in this figure are based
on polynomial interpolations between the numerical data.
Continuations of these graphs to larger values of $N$ ($N>27$ for 
$E_{\Delta S=1/2}$ and $N>36$ for $E_{\Delta S=1}$) are unwarranted.
}
\label{fig.3}
\end{figure}

\begin{table}[h]
\begin{center}	
\caption{Lowest energies per bond associated with the four 
different {\bf k}-points in the  Brillouin zone of the $N=36$ sample.}
\label{table 1}
\renewcommand{\arraystretch}{1.2}
\begin{tabular}{|c|c|}\hline\hline
$\hspace{1cm}{\bf k}\hspace{1cm}$       & \hspace{2cm}  E/N\hspace{2cm} \\ \hline \hline
$(0,0)$           &  -0.438377\\ 
$(\frac{2\pi}{3},0)$        &  -0.437851\\ 
$(0,\frac{2\pi}{\sqrt{3}})$  &  -0.437585\\ 
$(\frac{4\pi}{3},0)$        &  -0.438096\\  \hline \hline
\end{tabular}
\vspace*{5cm}  
\end{center}	
\end{table}

\begin{table}[h]
\caption[99]{
Expectation values of the chiral operator for triangular  
loops of various sizes:
d is the side-length of the triangle in units
of the nearest neighbor distance.
These expectation values have been computed
in the chiral ground states of the odd
samples ( they are homogeneous $ {\bf k} = (0,0)$ states
and belong to the
two degenerate complex conjugate IRs of $C_{3}$).
The lines with the entry (L.w-f) in the left column contain the expectation
values of the same operators in the variational Laughlin
wave-function of ref. \cite{ywg93}.
The last six lines contain the correlations between the chiralities 
on elementary triangular plaquettes for various distances.
}
\label{table 2}
\begin{tabular}{|cccc|} 
\hline\hline
             & \multicolumn{3}{c|}{ $<{\rm Im(P_{123})}>$ }         \\ \hline
 N           &     9          &       21       &        27       \\ 
 d=1          &   0.2570    &     0.0353       & 0.0567         \\
  (L.w-f)     &   0.2780    &     0.4022       & 0.4033         \\
 d=$\sqrt{3}$ &   0.0613    &     0.0261       & 0.0241         \\
  (L.w-f)     &   0.1176    &     0.0105       & 0.0081         \\
 d=2          &   0.0000    &     0.0094       & 0.0004         \\
  (L.w-f)     &   0.0000    &     0.0098       & 0.0071         \\
 d=$\sqrt{7}$ &             &     0.0013       & 0.0013         \\
  (L.w-f)     &             &     0.0035       & 0.0020         \\
 d=3          &	            &                  & 0.0080         \\
  (L.w-f)     &             &                  & 0.0015         \\
 d=$\sqrt{12}$&             &                  & 0.0000         \\
 \hline\hline
             & \multicolumn{3}{c|}{$<{\rm Im(P_{123})Im(P_{456})}> - <{\rm Im(P_{123})}>^{2} $ } \\ \hline
shell-shell distance	&	             &   N=21       &      N=27    \\
0-0          &  	   & 0.7013        &     0.6990       \\
0-1         &              &  -0.0012      &     0.0135      \\
0-2        &               &  -0.0166      &     0.0131	   \\
0-3         &              &  -0.0017      &     0.0019     \\
0-4          &             &   0.0072      &    -0.0000     \\
0-5        &               &               &    -0.0029      \\
\hline
\end{tabular}

\end{table}
\clearpage

%
%

\end{document}